\documentclass[a4paper,11pt]{article}
\pdfoutput=1 % if your are submitting a pdflatex (i.e. if you have
\usepackage{jinstpub} % for details on the use of the package, please
\title{\boldmath Magnetic field measurements on the  mini-ICAL detector using Hall
probes}

\author[a,b,c,1]{Honey,\note{Corresponding author.}}
\author[b]{B. Satyanarayana,}
\author[b]{R. Shinde,}
\author[a]{V.M. Datar,}
\author[a,c]{D. Indumathi,}
\author[b]{Ram K V Thulasi,}
\author[d]{N. Dalal,}
\author[d]{S. Prabhakar,}
\author[d]{S. Ajith,}
\author[d]{Sourabh Pathak,}
\author[d]{Sandip Patel}
\affiliation[a]{The Institute of Mathematical Sciences,\\ Taramani, Chennai 600113, India}
\affiliation[b]{Tata Institute of Fundamental Research,\\ Homi Bhabha Road,
Mumbai 400005, India}
\affiliation[c]{Homi Bhabha National Institute,\\ Anushakti Nagar, Mumbai 400094, India}
\affiliation[d]{Bhabha Atomic Research Centre, \\Mumbai 400085, India}

\emailAdd{honey@tifr.res.in}

\abstract{The magnetised 51 kton Iron Calorimeter (ICAL) detector proposed to be
built at INO is designed with a focus on detecting 1-20 GeV muons. The
magnetic field will enable the measurement of the momentum of the $\mu^{-}$ and $\mu^{+}$
generated from the charge current interactions of $\nu_{\mu}$ and
$\overline{\nu}_\mu$ separately within iron in the detector, thus permitting the
determination of the neutrino mass ordering/hierarchy, among other important 
goals of ICAL. Hence it is important to determine the
magnetic field as accurately as possible. The {\sf mini-ICAL} detector
is an 85-ton prototype of ICAL, which is operational at Madurai in South
India. We describe here the first measurement of the magnetic field in
{\sf mini-ICAL} using Hall sensor PCBs. A set-up developed to calibrate
the Hall probe sensors using an electromagnet. The readout system has
been designed using an Arduino Nano board for selection of channels of
Hall probes mounted on the PCB and to convert the analog voltage to a
digital output. The magnetic field has been measured in the small gaps
(provided for the purpose) between iron plates in the top layer of {\sf
mini-ICAL} as well as in the air just outside the detector. A precision
of better than 3\% was obtained, with a sensitivity down to about 0.03
kGauss when measuring the small fringe fields outside the detector.}


\begin{document}
\maketitle
\flushbottom

\section{Introduction and Motivation}
\label{sec:intro}
The proposed magnetized Iron Calorimeter (ICAL) detector to be located
at the India-based Neutrino Observatory (INO) is designed to be a 51
kton iron detector optimized to detect $\mu^{-}$ and $\mu^{+}$ generated
from the charged current interactions within iron of $\nu_{\mu}$ and
$\overline{\nu}_\mu$ respectively in the energy range of few to 10s
of GeV. The detector will comprise 3 modules made up of 151 layers of
56 mm thick iron plates with 150 layers of Resistive plate chambers
(RPCs) as active detector element which are sandwiched between the iron
plates. The magnetic field will be generated through current passing in
copper coils which are wound around the iron plates in slots constructed
for the purpose. Hence the field is expected to vary in both magnitude
and direction over the area of the iron plates.

One of the most important capabilities of ICAL will be its ability to
distinguish the sign of charged particles through the iron with a magnetic
field of 1--1.5 T (10--15 kGauss) that will be generated using copper
coils. Apart from helping to distinguish neutrino and anti-neutrino
induced events, the magnetic field is also crucial to reconstruct the
momenta (both magnitude and direction) of the muons produced in these
interactions which is a critical input for the corresponding physics
analyses \cite{ino:report},\cite{indu:madam}.

Simulations for the magnetic field using ICAL geometry have been performed
using MAGNET software \cite{sp:bhr}. In order to validate these
simulations and find the correlation between the estimated magnetic field
and the actual magnetic field, measurements are needed. Hence it is
important to precisely measure and calibrate the magnetic field in ICAL.

To understand the challenges in the construction of such a huge
detector, a small prototype detector called {\sf mini-ICAL} has been
set-up in Madurai, Tamil Nadu. The 85-ton detector, comprising 11 layers
of iron with planar dimensions of $4 \times 4$ m$^2$ in the $x$-$y$
plane, has been magnetised to about 1 T using a current of 500 A. In
each layer, the required geometry is achieved by the use of 7 iron
plates of slightly varying dimensions, with small (2--3 mm) gaps between
them. We
report here for the first time, results of a study of the magnetic field
generated in the {\sf mini-ICAL} in a limited region of the detector, in
particular, near these gaps where it is possible to insert appropriate
probes.

There are various techniques which are available to measure magnetic
fields, for example, SQUID, NMR, Magneto-resistance, Magneto-optical, Hall
sensors, etc \cite{st:students}. SQUIDs require very low temperature to
operate while NMR, Magneto-resistance and Magneto-optical devices cannot
be used due to their low range and scalar field measuring technique;
hence these are not found suitable for our study.  Hall sensors provide
reasonably good accuracy and can measure fields over a wide range
(100--10000 mT); therefore Hall sensors were chosen as the primary
sensor to study the magnetic field in the {\sf mini-ICAL} detector. In
particular, the field measurement probe used in the current study was done 
at low cost ($\sim$ US $\$$ 100). It was designed and constructed using 
inexpensive Hall sensors with sensitivity down to 10 mT and standard 
electronics. Since measurement and calibration of the magnetic
field in the 151 layers of ICAL (when built) is likely to be a recurring
activity, it was gratifying that the study showed consistency and could
accurately measure the field in the small gaps/spaces between different
iron plates in a layer. Even in regions such as in the air {\em just
outside} the detector, where the magnetic field is expected to be small,
of the order of 10s of Gauss, our measurements with different sensors
were consistent to within 15--20 Gauss, which is close to the least
count of the instrument.

We report here both on the magnetic field measurements in the air gap
between iron plates in the top layer of {\sf mini-ICAL} as well as just
outside one of these gaps, in the air outside the detector. Some
preliminary results were reported in Ref.~\cite{hk:honey}. The former is
expected to be close to the value of the magnetic field in the adjoining
iron layer. It will thus be possible to validate the estimates of the
magnetic field obtained by the simulations. An actual simulation study
of the {\sf mini-ICAL} detector has not yet been done and is thus outside
the scope of the present work.

We first describe the relevant geometry of the {\sf mini-ICAL} detector
before going on to describe the basic characteristics of Hall sensors,
calibration of Hall sensors using an electromagnet
and designing the Hall probe readout using Arduino Nano to extract the
signals from Hall sensors. Finally, we report the first measurement of
the magnetic field in the {\sf mini-ICAL} detector.

\section{The mini-ICAL detector}

The {\sf mini-ICAL} detector, see Fig.~\ref{miniICAL}, is an 85-ton
prototype magnetized detector located in Madurai, which detects cosmic
muons. A magnetic field of upto 1.5 T can be produced in it with a
current of 900 A, with upto 90\% of its volume having a magnetic field
of more than 1 T near saturation.  %

\subsection{The mini-ICAL geometry}

\begin{figure}[!tbp]
  \centering
  \begin{minipage}[b]{0.49\textwidth}
    \includegraphics[width=\textwidth]{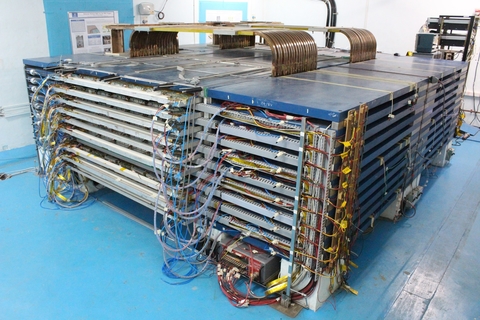}
    \caption{The {\sf mini-ICAL} detector with 11 iron layers. The
    upper portion of copper coils and slots are clearly visible.}
    \label{miniICAL}
  \end{minipage}
  \hfill
  \begin{minipage}[b]{0.49\textwidth}
    \includegraphics[width=\textwidth]{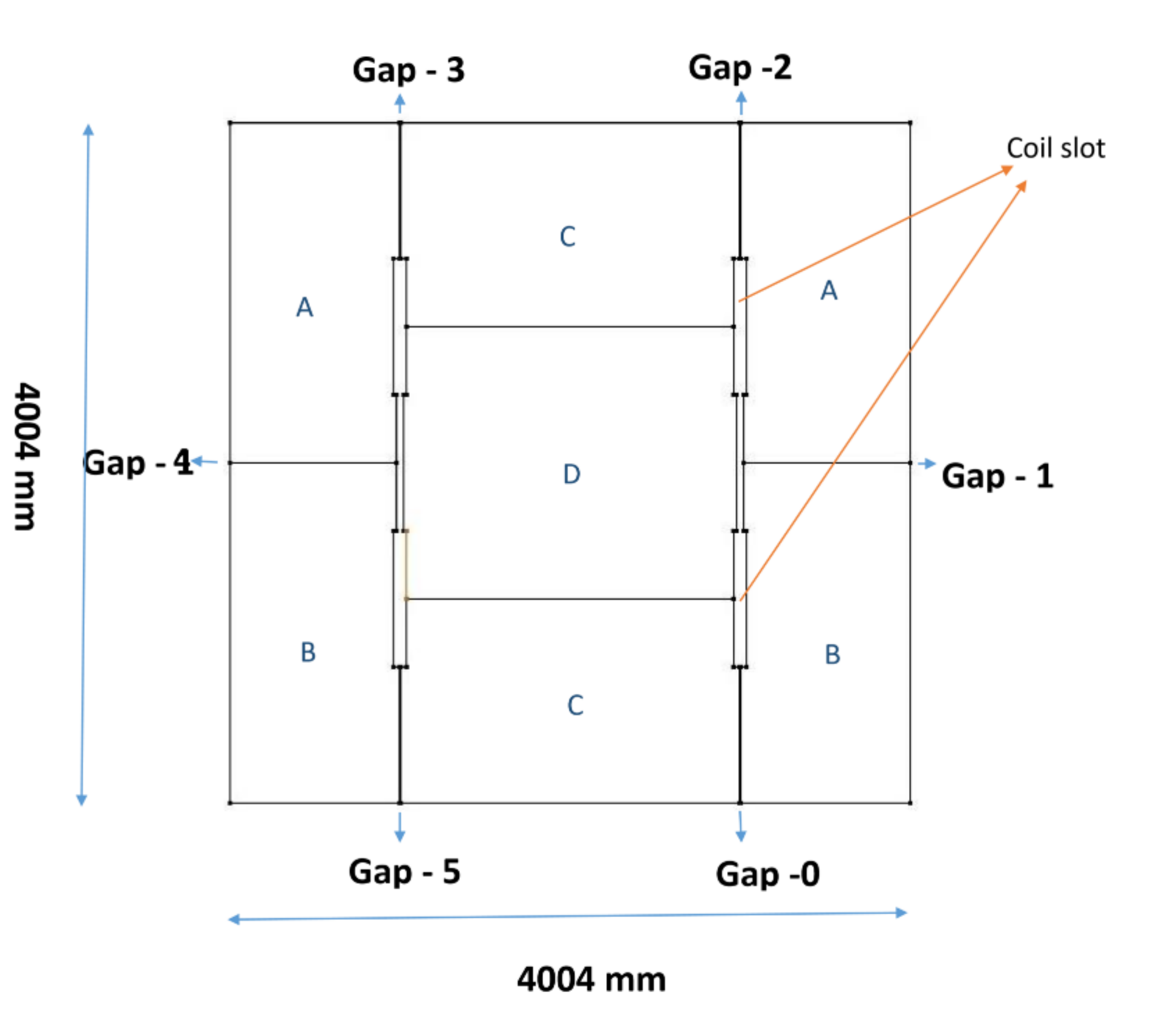}
    \caption{Schematic of top view of {\sf mini-ICAL} showing the gaps 0, 2, 3,
    and 5. The $B$-field is dominantly in the $x$ direction
    at these gaps.}
    \label{mini_ical_up}
  \end{minipage}
\end{figure}

The {\sf mini-ICAL} consists of 11 layers of 56 mm thick iron plates
with RPCs sandwiched in the 45 mm gap between them.  Each iron layer
is $4 \times 4$ m$^2$ in area and is assembled using 7 plates of iron
(2 A-type with dimensions 2001 $\times$ 1000 mm$^2$, 2 B-type with
dimensions 2001 $\times$ 1000 mm$^2$, 2 C-type with dimensions 2000 $\times$
1200 mm$^2$ and 1 D-type with dimensions 1962 $\times$ 1600 mm$^2$ as shown in
Fig.~\ref{mini_ical_up}).  The C-type plate is slightly narrower than
the A-type one, although all of them have the same long edge of 2 m. A
gap of about 3 mm, kept fixed using non-magnetic aluminium spacers, is
introduced between the iron plates in a layer in order to give access to
the (2 mm thick) Hall probe to measure the magnetic field.  This provision
for measuring the magnetic field measurement is available in only three
layers, the 1st (bottom), 6th (middle) and 11th (top) layers. The gap is
adjusted to be only 2 mm in the remaining layers\footnote{These gaps can
change when the magnetic field is turned on due to the attractive forces
between the magnetized plates.}. While the magnetic field measurement
can only be made in the gap between plates, this is expected to be close
to the value of the field in the adjacent iron plates. Gaps 0, 2, 3,
and 5 (see Fig.~\ref{mini_ical_up}) are where measurements of the magnetic
field have been made. Gaps 1 and 4 are not easy to access.

A current of 500 A is passed through two set of hollow copper coils, located in the
coil slots (see Fig.~\ref{mini_ical_up}), each having 18 turns, with area
of cross section $30 \times 30$ mm$^2$. Low conductivity ($<$ 10 micro siemens per cm)
 cooling water is flowed inside the copper coil for cooling purposes. \\

\subsection{Construction of the Hall probe}

\begin{figure}
\includegraphics[width=\textwidth]{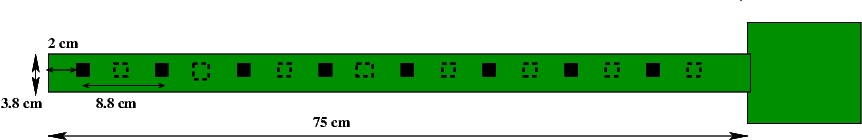}
\caption{Schematic of the Hall probe, showing the placing of the sensors
and the relevant dimensions.}
\label{hall_pc_dim}
\end{figure}

\begin{figure}
  \includegraphics[width=\linewidth]{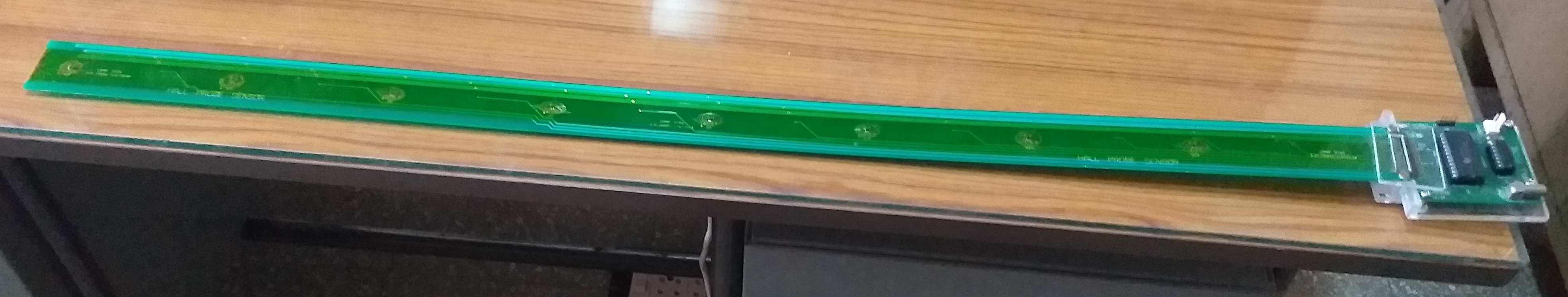}
  \caption{Hall probe PCB with Hall sensors mounted on it.}
\label{hall_PCB}
\end{figure}

The probes can only be inserted in the air gaps of about 3
mm width. From the literature survey, CYSJ106C GaAs Hall Effect Element
sensor \cite{data:sheet} was found to be suitable for the requirement
of Hall probe PCB. The specifications of the sensor are mentioned in
Table \ref{tab:label1}.  The probe is constructed by mounting 16 Hall
sensors labelled 0 to 15, of size 1.5 mm $\times$ 1.5mm $\times $ 0.8 mm
(thickness) on a PCB of length 75 cm about 4.4 cm apart, and on alternate
sides of it, as shown in Fig.~\ref{hall_pc_dim}; an actual picture is
shown in Fig.~\ref{hall_PCB}. The thickness of the PCB is 1.2 mm and
after mounting the Hall sensors the thickness of PCB becomes 2.3 mm. The
sensors are mounted alternatively on the PCB along its length. Therefore
when the PCB is placed in a uniform uni-directional field, according
to the direction of $B$-field that a particular sensor faces, half the
sensors will show positive value of Hall voltage and other half (mounted
on the other side of PCB) will show negative value of Hall voltage. Each
sensor is given a number according to their position from the front end
electronics. The sensor at the far end from the front end electronics is
numbered 0 and the sensor at the near end to the front end electronics
is numbered 15.\\

\begin{center}
\begin{table}[htp]
\begin{tabular}{|c| c| c|}  \hline
%  \Xhline{1\arrayrulewidth}
%  \Xhline{1\arrayrulewidth}
 \textbf {Parameter} & \textbf {Test Condition} & \textbf { Value} \\ [0.5ex] 
%  \Xhline{1\arrayrulewidth}
 \hline
Measurement Range    &  & 0--3 T   \\ \hline
Max. input power  &  & 150 mW    \\ \hline
Max input current$/$voltage &   & 13 mA/10 V \\ \hline
Operating temperature range  &  & $-40 \sim 125^\circ$ C    \\ \hline
\hline
 Hall output voltage & B = 100mT, Ic = 8mA, Vc = 6V & 110$\sim$150 mV \\
 \hline
 Offset voltage & Vc = 6V, B = 0mT & $\pm$ 11 mV \\
 \hline
  Input resistance & B = 0mT, Ic = 0.1mA & 650$\sim$850 $\Omega$ \\ 
 \hline
  Output resistance & B = 0mT, Ic = 0.1mA & 650$\sim$850 $\Omega$ \\ 
 \hline
  \shortstack{Temperature coefficient of \\ Hall output voltage}   & Ic = 5mA, B = 100mT & -0.06$\%$/$^0$C \\
 \hline
 \shortstack{Temperature coefficient of \\ Hall input and output resistanc}   & Ic = 0.1mA, B = 0mT & 0.3$\%$/$^0$C \\
\hline
Linearity & Ic = 5mA, B = 0.1/0.5T & 2$\%$ \\ \hline % [0.5ex] 
% \Xhline{2\arrayrulewidth}
\end{tabular}
% \vspace{3mm}
\caption{General characteristics and electrical specifications of the
CYSJ106C GaAs Hall sensor \cite{data:sheet}.} 
\label{tab:label1}
\end{table}
\end{center}

\subsection{Experimental setup}

\begin{figure}[bth]
   \centering
  \includegraphics[width=0.7\textwidth]{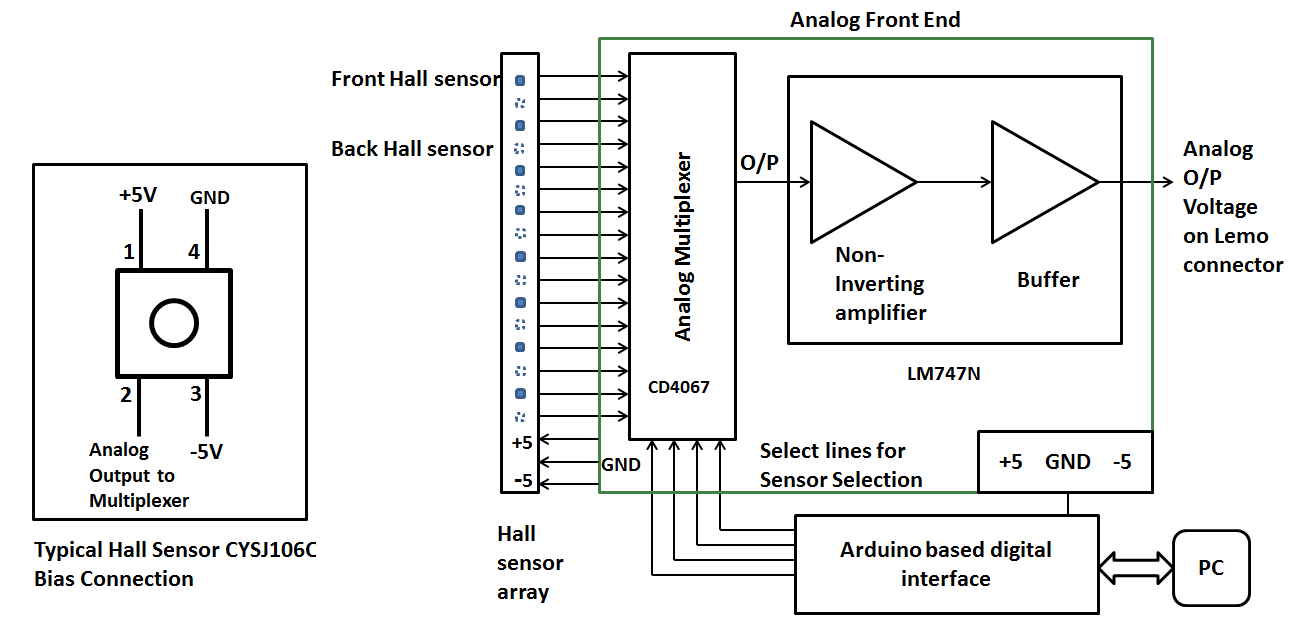}
  \caption{Schematic showing the set up for the Hall probe with the mounted sensors
  on the left side, the analog front end at the top, and the Arduino
  digital interface that connects to the PC shown at the bottom.}
  \label{adc_setup}
\end{figure}

The schematic circuit is shown in Fig.~\ref{adc_setup}.  Dual DC power
supply of $\pm$ 5 V has been used to bias the Hall sensors. All 16 analog
Hall voltages are available to read at a time, which are interfaced to
the small Analog front end board. The analog front end board consists
of a multiplexer, amplifier and buffer. For the sake of simplicity, 16:1
multiplexer has been introduced to the 16 analog readout channels one by
one. The IC CD4067-16 analog multiplexer with 4 digital select lines has
been used. This IC features low leakage current, lower power consumption
and low cross talk between the channels. The output of the multiplexer
is connected to a non-inverting amplifier with a gain of 1.045 followed
by unity gain amplifier for impedance matching to avoid loading of
the output stage. The dual operational amplifier IC LM747N is used to
configure the amplifier and buffer. This chip has low power consumption
and short-circuit protection. The output is available on a Lemo connector,
which can be read out using a digital multimeter or can be given to an
ADC with higher resolution. An Arduino Nano board is used to select
the Hall sensor channel during manual readout using a digital multimeter.

\section{Calibration of the Hall sensors}
\label{sec:calib}
There are two components to the calibration: one is to determine the
offset voltage (the reading of the Hall sensor in the absence of any
magnetic field), and the other is to calibrate the Hall voltage to
(a known) magnetic field.
\subsection{Measurement of the Offset Voltage}

\begin{figure}[bhp]
  \centering
  \includegraphics[angle=0,width=0.7\textwidth]{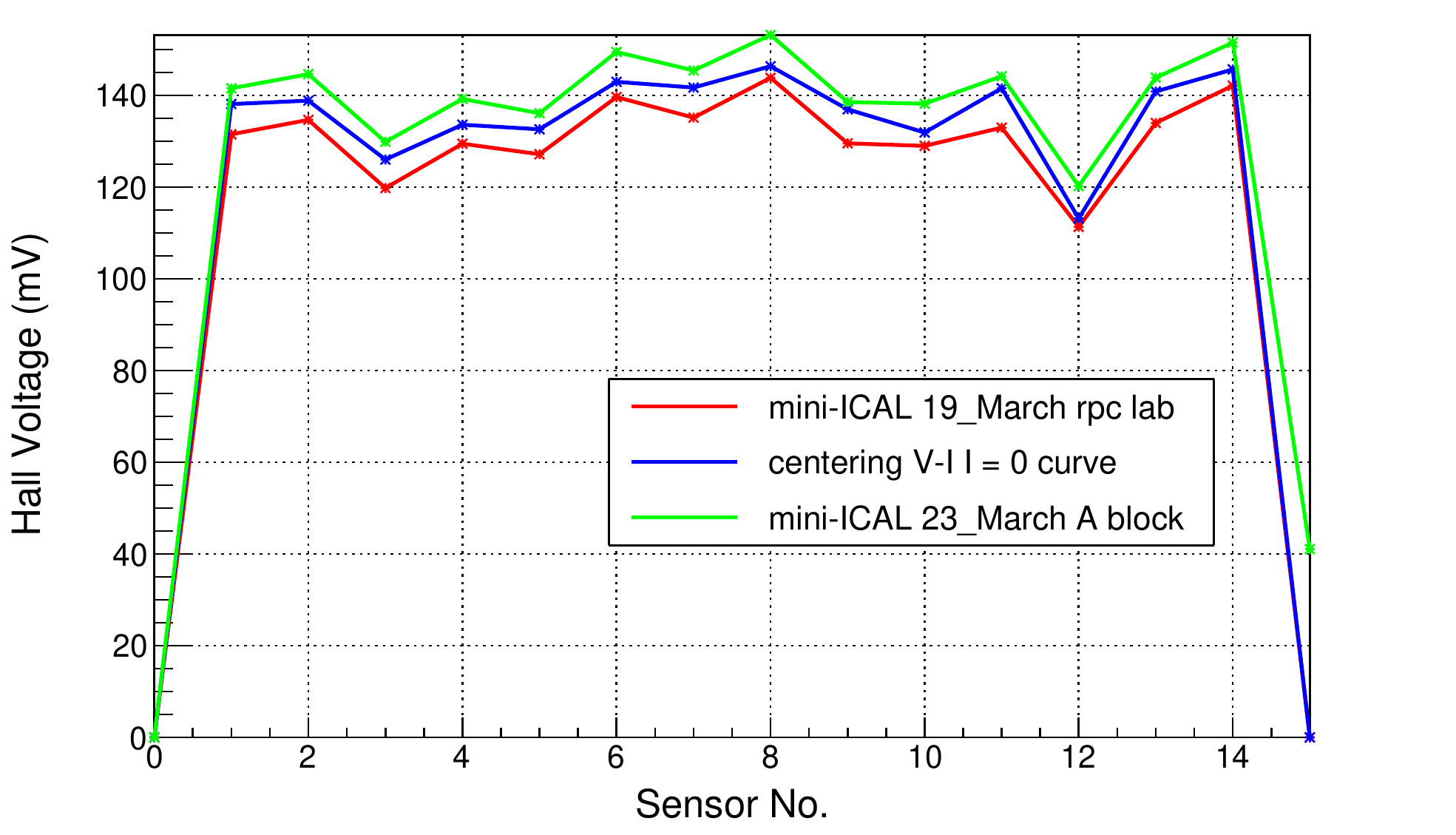}
  \caption{Offset voltage $V_0$ for each the Hall sensors measured away from
  {\sf mini-ICAL} at different places and times (in red and green) in
  comparison with the offset obtained by centering the $V$-$I$ curve (in
  blue); see text for details.}
  \label{offset_1}
\end{figure}

The offset voltage arises from the intrinsic properties of the sensor (see
Table~\ref{tab:label1}) as well as from the associated electronics. The
offset voltage for each sensor is measured in two different ways. In the
first approach, the offset of each Hall sensor is measured\footnote{Since
sensors 0 and 15 were defective, results are shown for sensors 1 to
14.} keeping the Hall sensor PCB away from the {\sf mini-ICAL} so that
only small stray magnetic fields are present (such as Earth's magnetic
field). These offset measurements were repeated at different times
and places. It is observed from Fig.~\ref{offset_1} where two typical
measurements are shown that there is a small variation in the measured
values of the offset.

\begin{figure}[!tbp]
  \centering
  \begin{minipage}[b]{0.49\textwidth}
    \includegraphics[width=7.2cm, height=4.8cm]{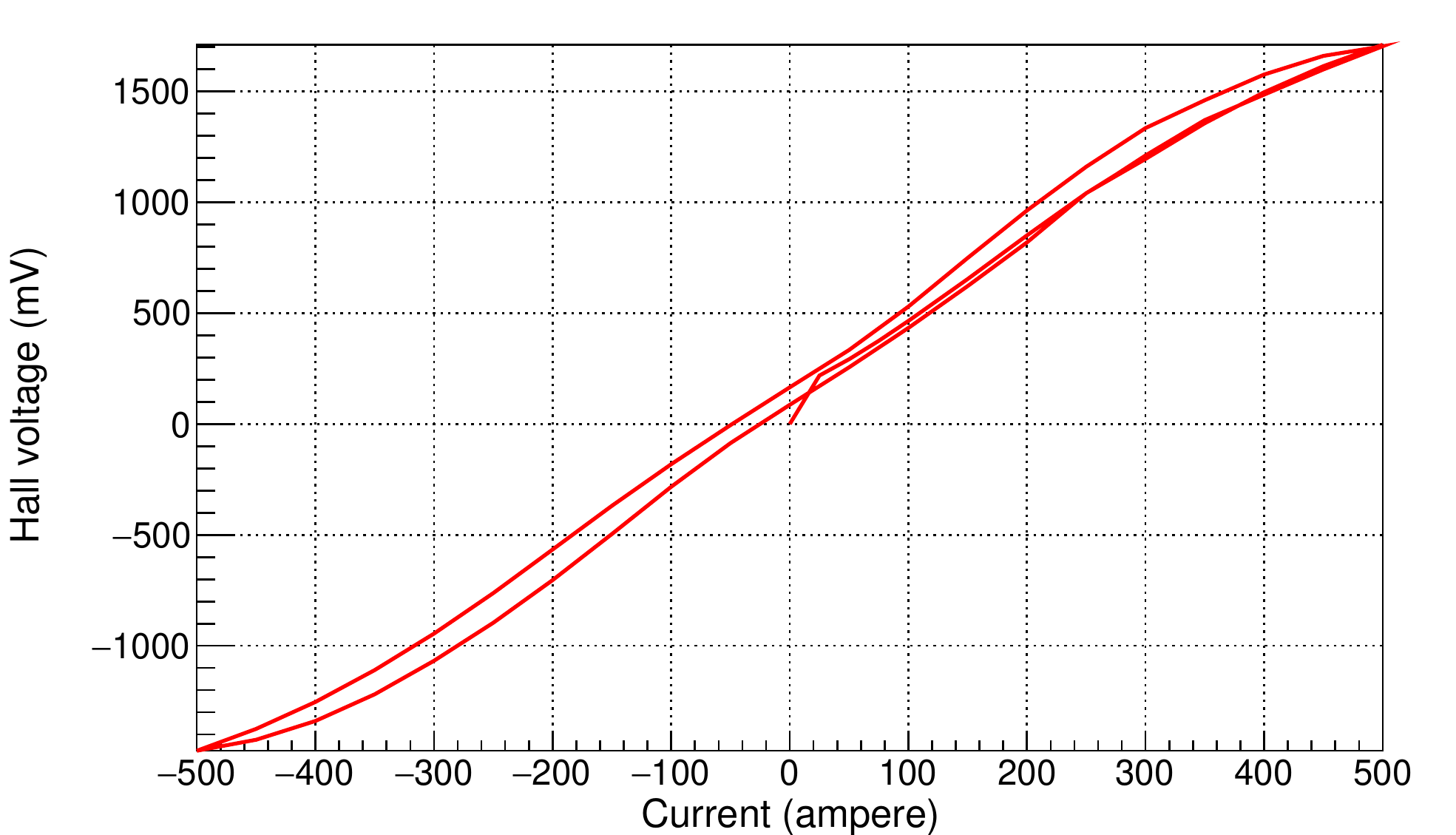}
    \caption{V-I curve for sensor 3 without offset correction.}
    \label{V_I_no_cor}
  \end{minipage}
  \hfill
  \begin{minipage}[b]{0.49\textwidth}
    \includegraphics[width=7.2cm, height=4.8cm]{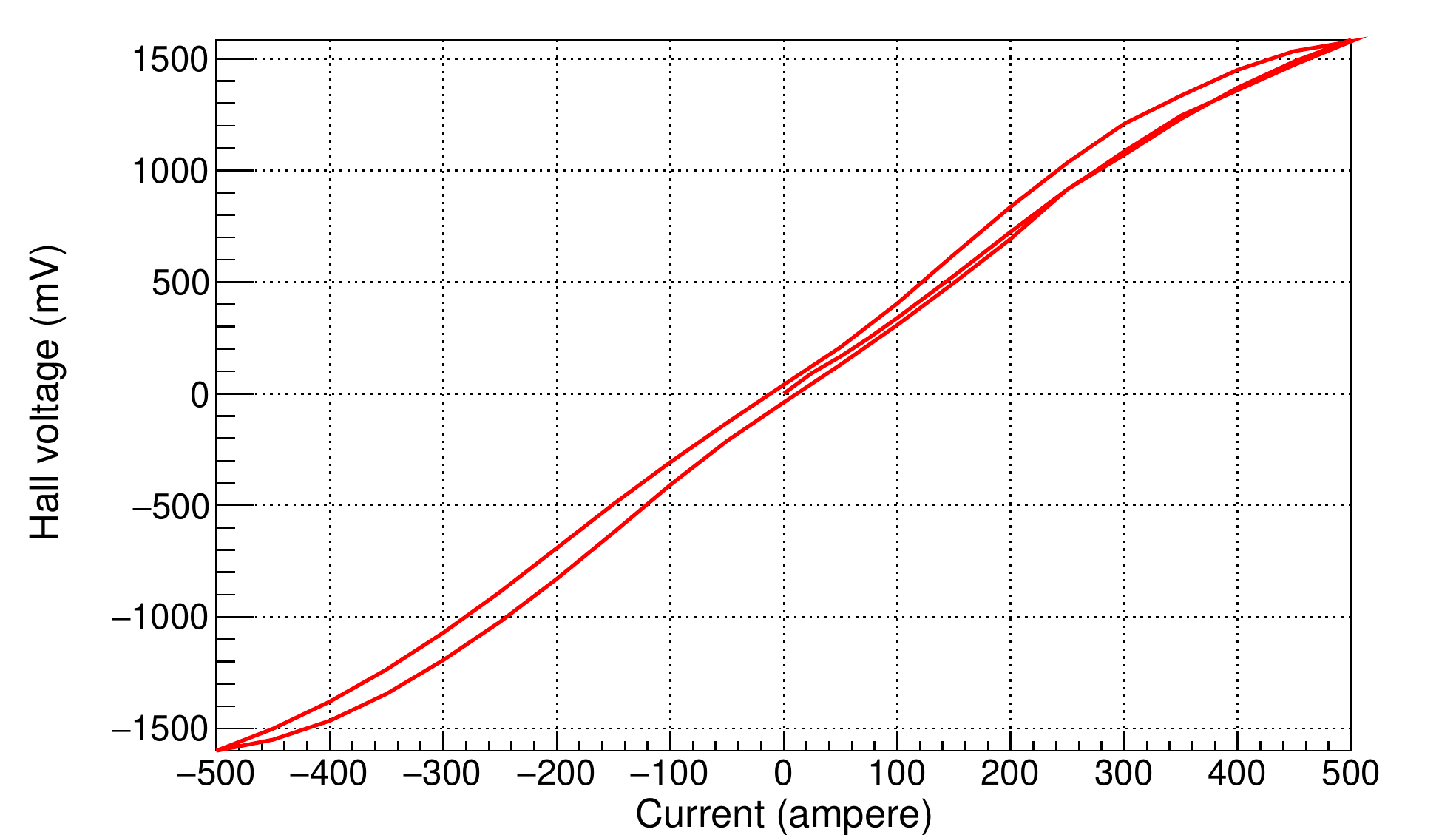}
    \caption{V-I curve for sensor 3 after offset correction.}
    % The $y$-axis on the right hand side shows the corresponding magnetic
    %field obtained after calibration; see next section.} 
    \label{V_I_cor}
  \end{minipage}
\end{figure}

The other way of finding the offset that is implemented here is by centering
the $V$-$I$ curve. The Hall probe was inserted into Gap-2 to the maximum
extent, so that Sensor 1 (15) was farthest from (closest to) the edge.
The voltage was measured for different values of current over an entire
hysteresis loop. It is seen from Fig.~\ref{V_I_no_cor} that the curve
is shifted upward so that an offset V$_{0}$ has to be subtracted in
order to center the $V$-$I$ loop; see Fig.~\ref{V_I_cor}. This offset was
measured for each Hall sensor and is shown in Fig.~\ref{offset_1} along
with the other measurements of the offset voltage. It is seen that the
offset measured in the two different ways follow the same pattern and
are consistent with one another. The variation between the measurements
of the offset voltage is estimated to be $\Delta V_0 = \pm 5$ mV, and
is later used to estimate the errors in the analysis.

Since the Hall sensors are alternatively mounted on opposite sides
of the PCB, one set of sensors (even/odd) sees a positive B-field and
other set of sensors (odd/even) sees a negative B-field. It was observed
that an extra offset V$_{\epsilon} = 8$ mV has to be added to all the
sensors to get agreement between the odd and even set of sensors. This
extra offset will be seen to amount to a field of about 0.05 kGauss.

%\section{Calibration of the Hall sensors}
%\label{sec:calib}

After subtracting the offset voltage, the Hall voltage measured by each
sensor is to be converted to a magnetic field measurement. This requires
calibration of the sensors by applying a known magnetic field.

\subsection{Calibration using gap dipole electro-magnet}
\begin{figure}
  \centering
\includegraphics[width=0.7\textwidth]{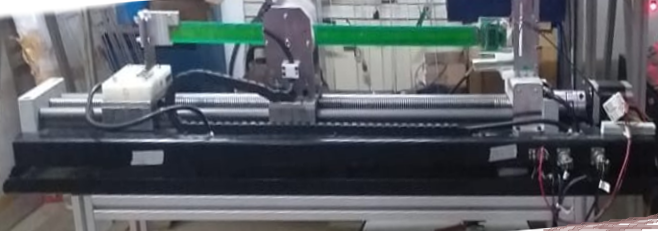}
  \caption{Calibration Instrument set-up.}
  \label{calibration_instrument}
\end{figure}

\begin{figure}
  \centering
\includegraphics[width=0.7\textwidth]{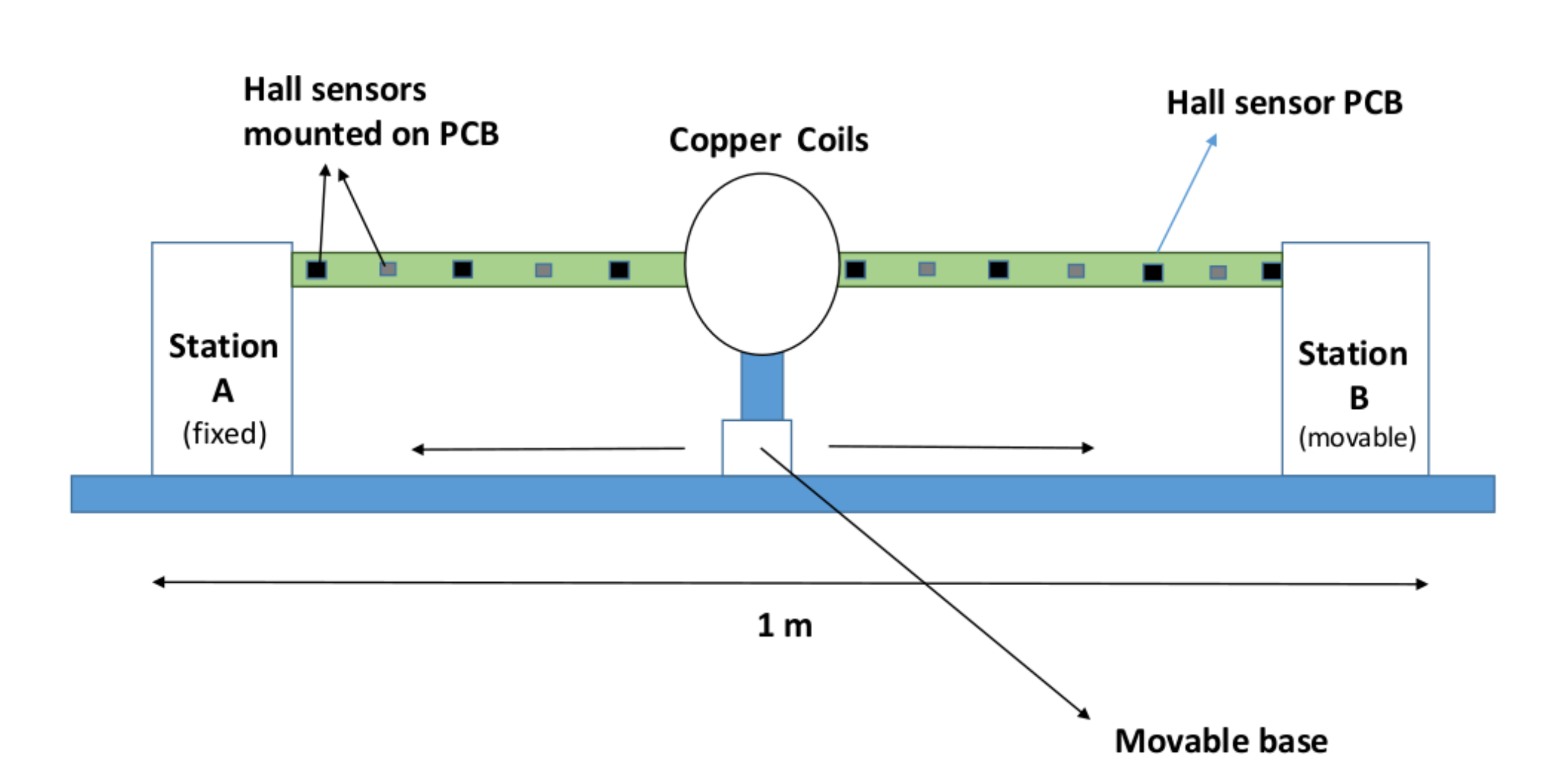}
  \caption{Block diagram of Calibration Instrument set-up.}
%  \label{block_diagram calibration_instrument}
\label{calib_block}
\end{figure}

 An electromagnet calibrator unit that can generate magnetic fields
upto 1.5 T (15 kGauss) by varying the current in the coil, is used. The
calibration set-up shown in Fig.~\ref{calibration_instrument} consists
of a set of two copper coils 14 mm in diameter which are mounted on a
base. The base can be moved forward and backward on a 1 m length platform
using a stepper motor. There are two stoppers named as
station 'A' (fixed) and 'B' (moveable) at each end of the platform; see
Fig.~\ref{calib_block}. Clamps are mounted on these two stoppers to
hold the Hall probe PCB in such a way that a given Hall sensor will
be positioned at the center of the poles of the coil which are 2.5 mm
apart. This will ensure the sensor sees a uniform and perpendicular
field generated by the coils.

\begin{figure}[!tbp]
  \centering
  \begin{minipage}[b]{0.4\textwidth}
    \includegraphics[width=6.5cm, height=4.8cm]{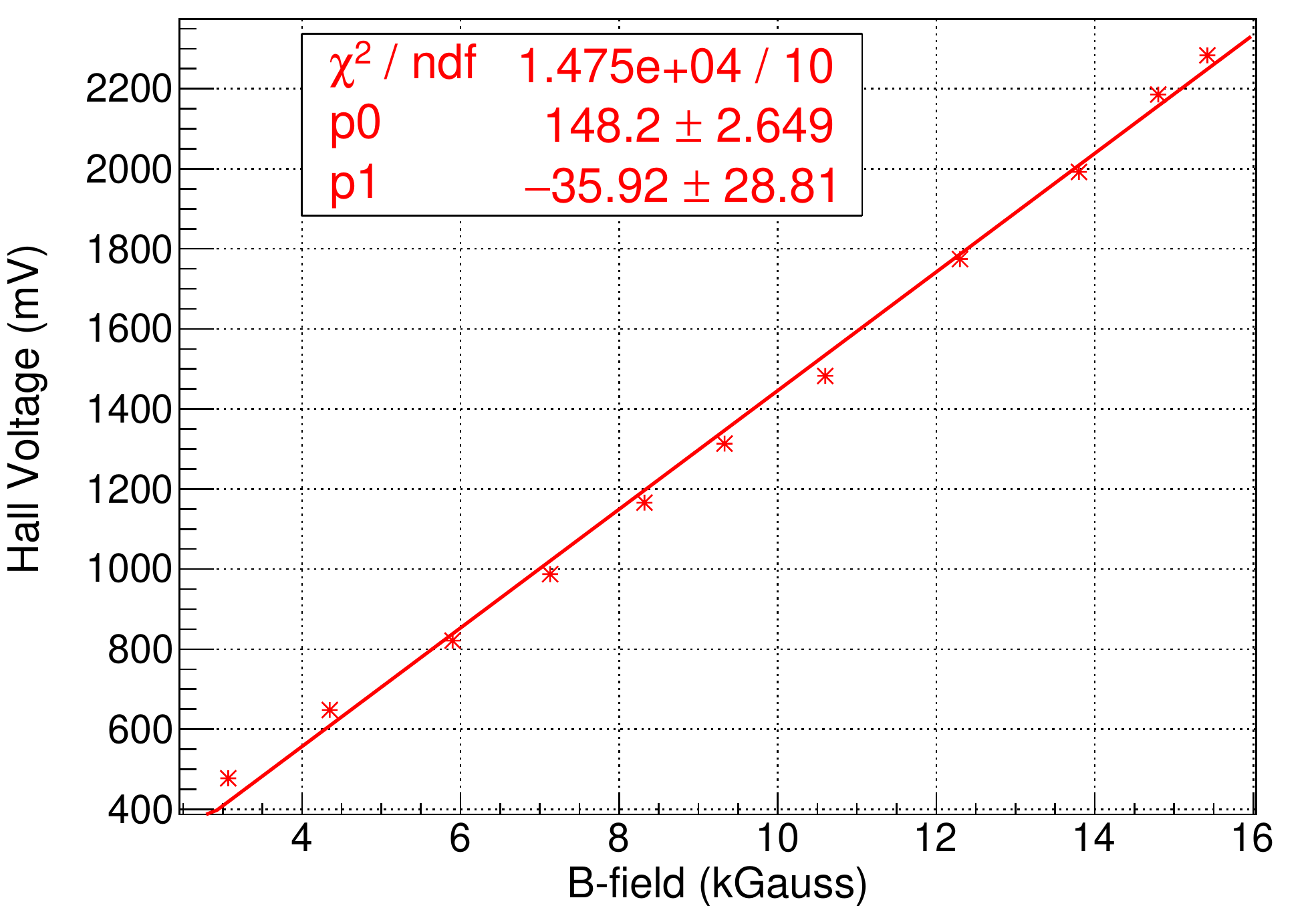}
    \caption{Calibration plot of sensor 2 for positive output voltage.}
    \label{cal_pos}
  \end{minipage}
  \hfill
  \hspace{-5mm}
  \begin{minipage}[b]{0.4\textwidth}
    \includegraphics[width=6.5cm, height=4.8cm]{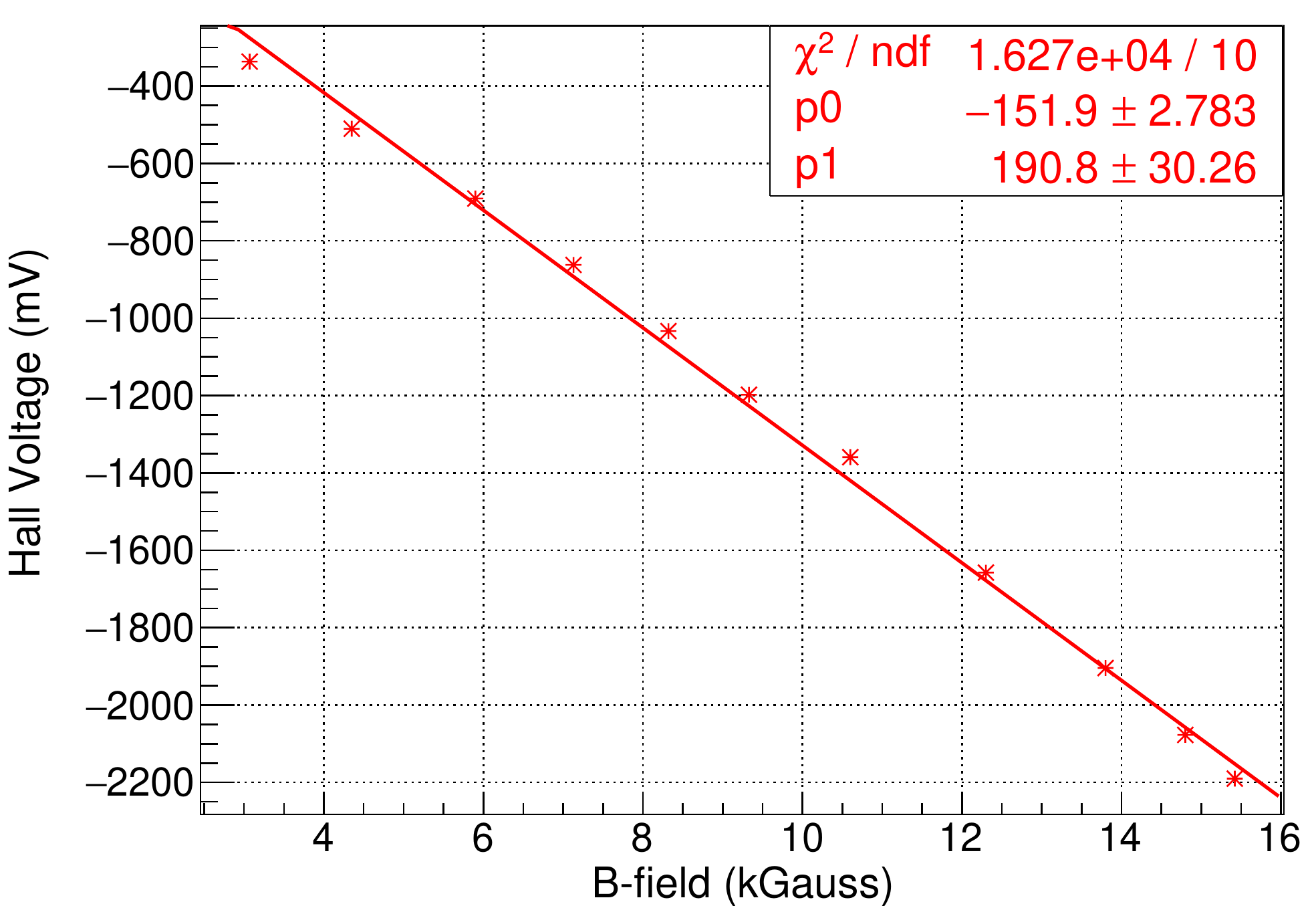}
    \caption{Calibration plot of sensor 3 for negative output voltage.}
    \label{cal_neg}
  \end{minipage}
\end{figure}

The B-field was varied by using different values of current supplied to
the coils by a DC regulated power supply
(model SVL 030010 D manufactured by the Sairush Electronic Systems).
Power supply is used in CC mode to get constant output current from 1
to 4 A in steps of 0.5 A and from 4 to 8 A in steps of 1 A.
The B-field generated between the coils is measured using standard gauss
meter at different values of the current supplied to the coils. The
corresponding Hall voltage of the Hall probe sensor is recorded. This
procedure was repeated for each sensor for both ramp-up and ramp-down
values of the current. The voltage obtained is plotted against the known
magnetic field. A linear relationship was found, as can be seen from the
Figs.~\ref{cal_pos} and \ref{cal_neg} plotted for sensor 2 and 7 respectively and the
slope $m$ is determined. This was repeated for each sensor. It was found that
the typical error on the fit was $\Delta m = \pm 3$ mV/kGauss.

The slopes and intercepts (latter not used in the analysis) for all the
Hall sensors are listed in Table~\ref{tab:label2} for both positive and
negative Hall voltages (differing directions of magnetic field). The
slope $m$ can then be used to calculate the magnetic field from the measured
voltage $V$ as, $B = (V- V_0+V_\epsilon)/m$.

\begin{table}[htp]
\centering
\begin{tabular}{|c|r|r|r|r|} 
 \hline
 & \multicolumn{2}{|c|}{\bf Side I} & \multicolumn{2}{|c|}{\bf Side II} \\ \hline
 \textbf{Sensor no.} & \textbf{Slope} & \textbf{Intercept} & \textbf{Slope}& \textbf{Intercept}\\  
 & (mV/kGauss) & (mV) & (mV/kGauss) & (mV) \\ \hline
% \hline
%0   &         &                 &            &           \\

1   &   -150.2    &     194.7   &    151.1   &      -38.19      \\
\hline
2   &    148.2    &    -35.92   &   -150.3   &       207.4      \\
\hline
3   &   -151.9    &    190.8    &    151.4   &      -46.74      \\
\hline
4   &    147.5    &    -35.38   &   -150.9   &       203.1      \\
\hline
5   &   -147.2    &     188.8   &    151.2   &      -52.18      \\
\hline
6   &    148.1    &    -33.38   &   -150.6   &        214.0              \\
\hline
7   &   -149.8    &     217.1   &    150.9   &     -43.7           \\
\hline
8   &    148.7    &    -26.35   &   -151.1   &     227.9             \\
\hline
9   &   -149.1    &     199.7   &    151.1   &     -55.73             \\
\hline
10  &    147.7    &    -19.88   &   -150.5   &     209.2              \\
\hline
11  &   -149.3    &     222.3   &    149.7   &    -40.59             \\
\hline
12  &    149.2    &    -49.86   &   -147.6   &      175.8            \\
\hline
13  &   -151.4    &     215.5   &    148.5   &      -31.91           \\
\hline
14  &    150.3    &    -27.16   &    -149.2  &        212.0           \\
\hline
%15  &             &             &            &               \\ 
% \hline
\end{tabular}
\vspace{2mm}
\caption{Calibration of Hall voltage to magnetic field: the slope and
intercept corresponding to each sensor for both positive and negative
field values. Side I corresponds to odd sensors facing a negative field
while Side II corresponds to odd sensors facing a positive field. The
even sensors always see the opposite field.}
\label{tab:label2}
\end{table}

%------------------------------cal plots--------------------------------------

\section{Error Estimation}
Before we go on to the actual measurement of the magnetic field in
{\sf mini-ICAL} we make a note of the errors involved. The sources of error
we have considered are in the measurement of the offset voltage, $\Delta
V_{0}$ and in the calibration fits to the slope, $\Delta m$. Hence the
magnetic field $B$ corresponding to a measured voltage of $V$ (and its
error $\delta B$) is given by
\begin{eqnarray} \nonumber
B  & = & \frac{V-V_{0}+V_{\epsilon}}{m}~, \nonumber \\
\delta {B} & = & \frac{B}{m} \sqrt{\frac{\Delta {V_{0}}^2}{B^2} +\Delta m^2}~.
\label{eq:formula}
\end{eqnarray}
It can be seen that the error in the measurement of the slope dominates
at large values of field while the error in the measurement of the
offset voltage dominates at small field values.

\section{Measurement of the magnetic field in  mini-ICAL}

Having completed the calibration of the Hall sensors, we then used them
to measure the magnetic field in {\sf mini-ICAL} The B-field was measured in
the gaps numbered 0, 2, 3 and 5 (see Fig.~\ref{mini_ical_up} for reference)
in the top layer of {\sf mini-ICAL} by inserting the Hall probe PCB into these
gaps; recall that sensor 1 is closest to the copper coil while sensor
15 is closest to the outside edge of the detector.

% Done till here

\begin{figure}[!tbp]
  \centering
  \begin{minipage}[b]{0.49\textwidth}
    \includegraphics[angle=0,width=\textwidth]{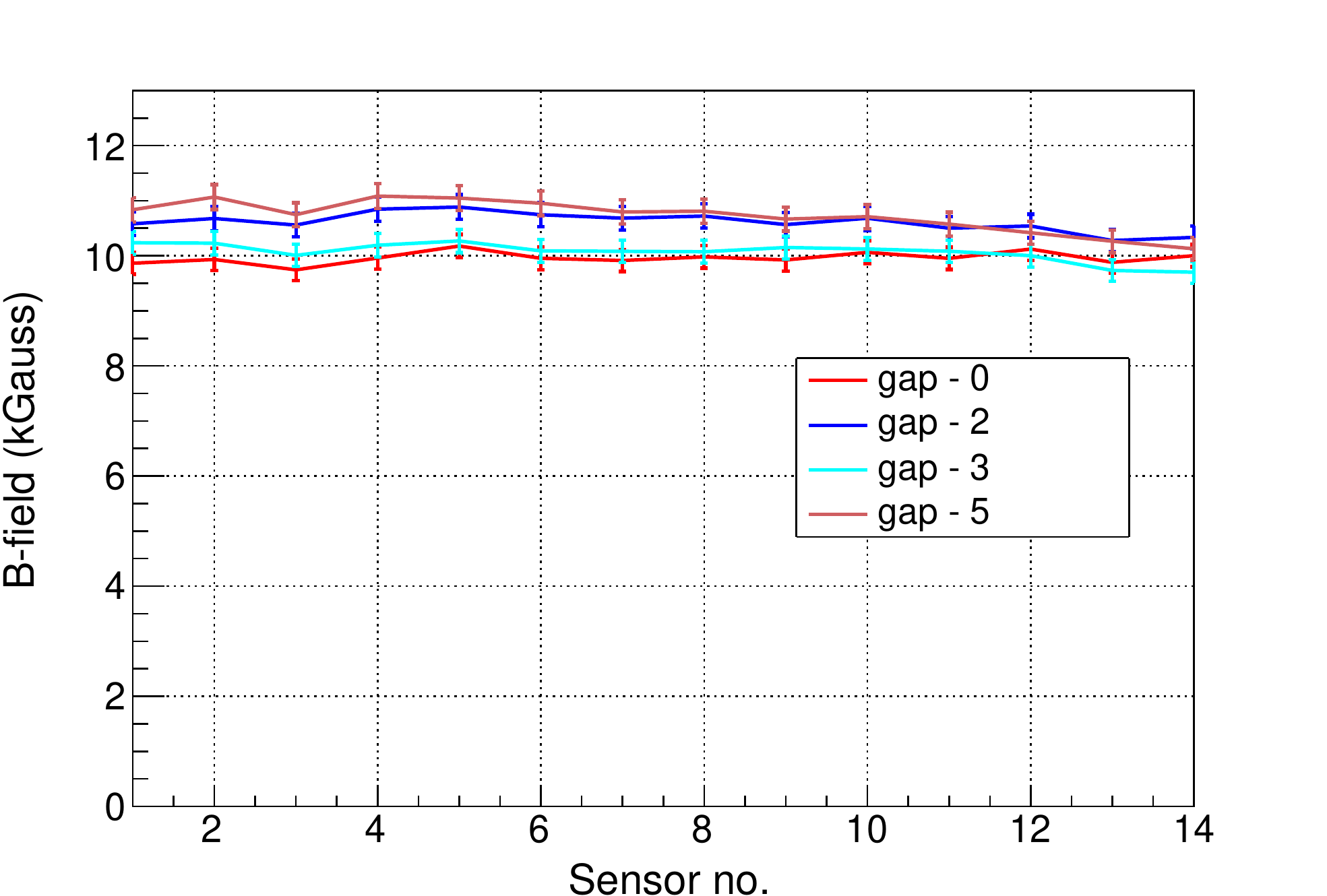}
    \caption{$B$-field measured in Gaps 0, 2, 3 and 5 by sensors 1--14.
    Note that sensor 14 typically measures a smaller field due to
    its location near the outer edge of the detector.}
    \label{gap_field}
  \end{minipage}
  \hfill
  \begin{minipage}[b]{0.49\textwidth}
    \includegraphics[angle=0,width=\textwidth]{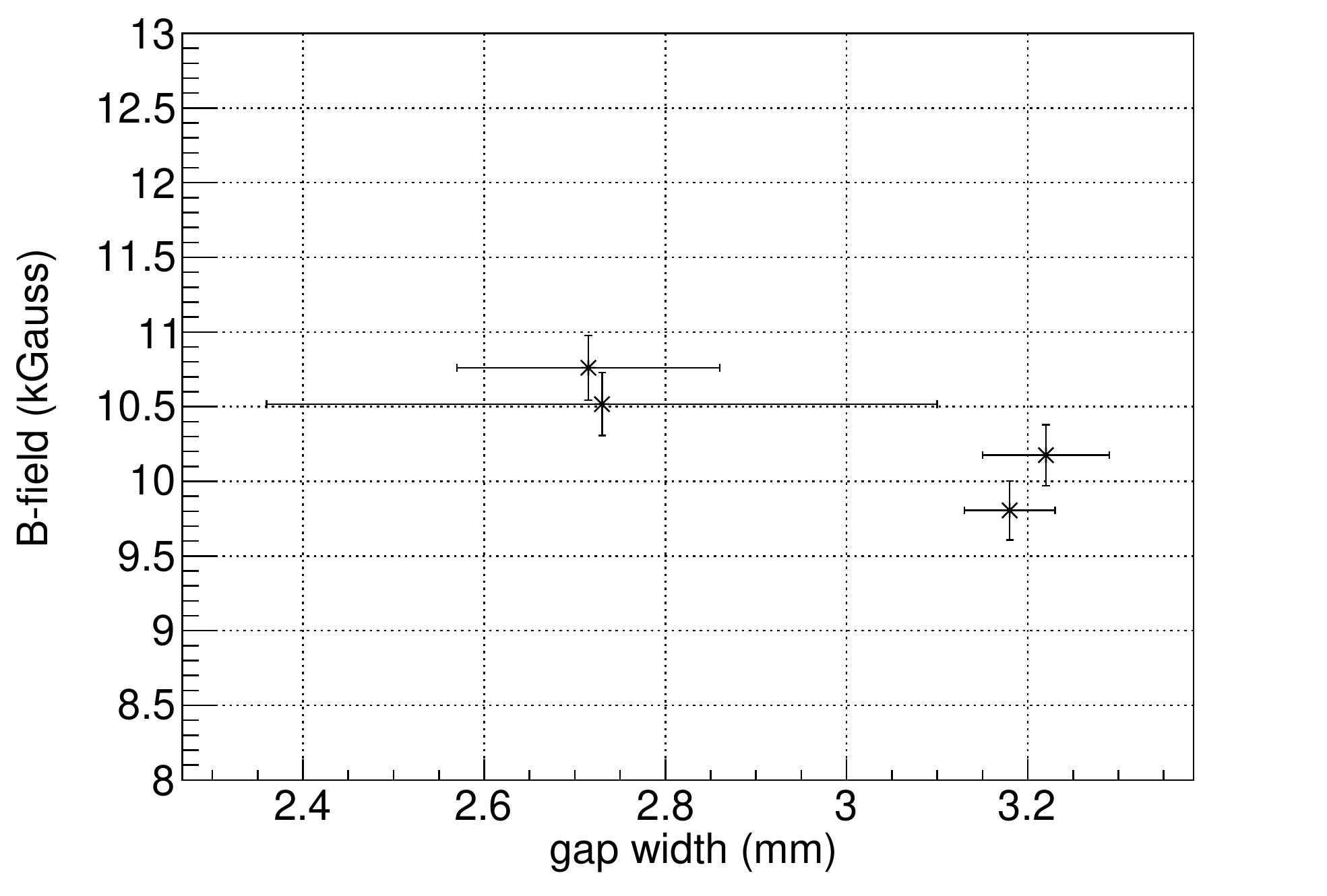}
    \caption{The figure shows the variation of the $B$-field as a
    function of the gap width for sensor 9. The field decreases
    perceptibly as the width increases.}
    \label{gap_vs_field}
  \end{minipage}
\end{figure}

The {\sf mini-ICAL} magnet is ramped up to 500 A using a DC power supply. The
Hall voltage measured from each sensor is determined, from which the
corresponding magnetic field is calculated as per Fig.~\ref{offset_1}
and Table~\ref{tab:label2}. This is shown in Fig.~\ref{gap_field}. Sensor
1 is closer to the copper coil while sensor 14 is closer to the edge and
hence the later measures a consistently smaller field in all gaps.

As can be seen from the geometry in Fig.~\ref{mini_ical_up}, the gaps 0,
2, 3 and 5 are symmetric and ideally should have the same magnetic field
(in the $x$ direction, upto a sign). However, in the actual assembly
of {\sf mini-ICAL} plates, these gaps vary in width, with the range shown in
Table~\ref{tab:label3}. Therefore the magnetic field is different for
these four gaps; the variation in the B-field for sensor 9 is shown as
a function of the mean gap width in Fig.~\ref{gap_vs_field}; as
expected, the field value decreases as the gap width increases.
Estimating the true value of the magnetic field in the iron plates from
the field values in the gap therefore requires detailed simulation,
although it is expected to be close to the field value in the gap. This
work is on-going and is beyond the scope of the present analysis.

\begin{table}[htp]
\centering
\begin{tabular}{|c|c|c|} 
 \hline
  \textbf{Gap no.} & Gap width (mm) & Avg width (mm) \\  
\hline
0   &  3.13-3.23   &  3.18\\
\hline
2  &   2.36-3.10   &  2.73  \\
\hline
3  &    3.15-3.29   &  3.22\\
\hline
5  &    2.57-2.86   &  2.715 \\ 
 \hline
\end{tabular}
\vspace{3mm}
\caption{Measured widths of gaps for Gaps 0, 2, 3, and 5 showing both
the range of values and their average.}
\label{tab:label3}
\end{table}

\section{Fringe-field just outside  mini-ICAL}

The fringe field (field outside the iron) is important because it is an
indicator of the amount of field leaking out of the detector. The fringe
field was measured by inserting the Hall probe into Gap 2 and sliding
the probe out by 1 cm at a time upto a distance of $L=70$ cm. Each
sensor measures the magnetic field in its vicinity. Since sensor
14 is closest to the edge, the magnetic field value it measures
drops the fastest as it exits the detector into the air, while
the magnetic field barely begins to drop for $L=70$ cm for sensor 1.
Hence for convenience, the probe geometry (given in
Fig.~\ref{hall_pc_dim}) was used to rescale the sensor position to
reflect the distance $x$ from the edge of the detector. Negative values
of $x$ indicate that the sensor is still inside the detector dimensions
while $x > 0$ indicates the sensor is making measurements in the air
outside the detector.

\begin{figure}[!tbp]
  \centering
\includegraphics[angle=0,width=0.49\textwidth]{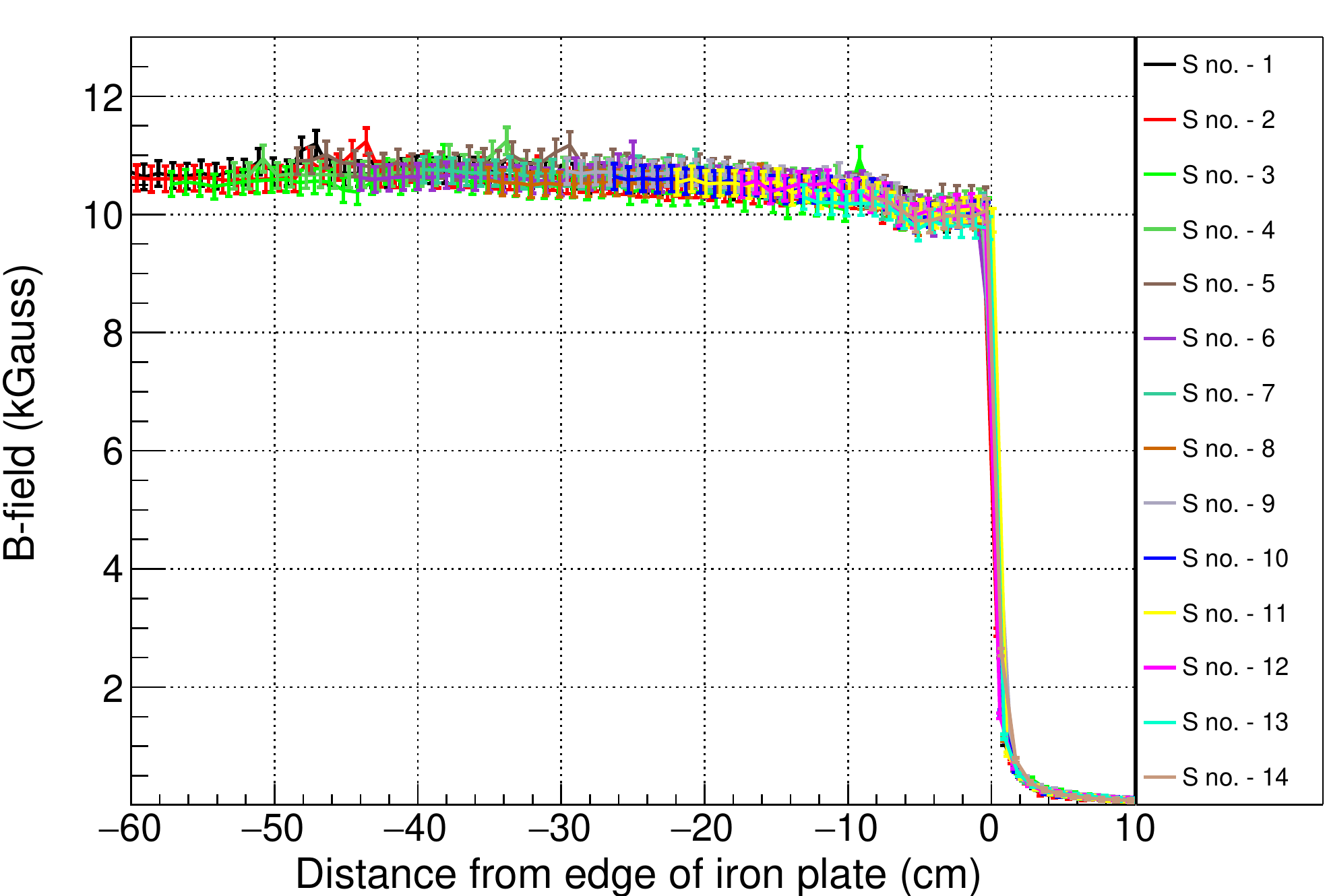}
\includegraphics[width=0.49\textwidth]{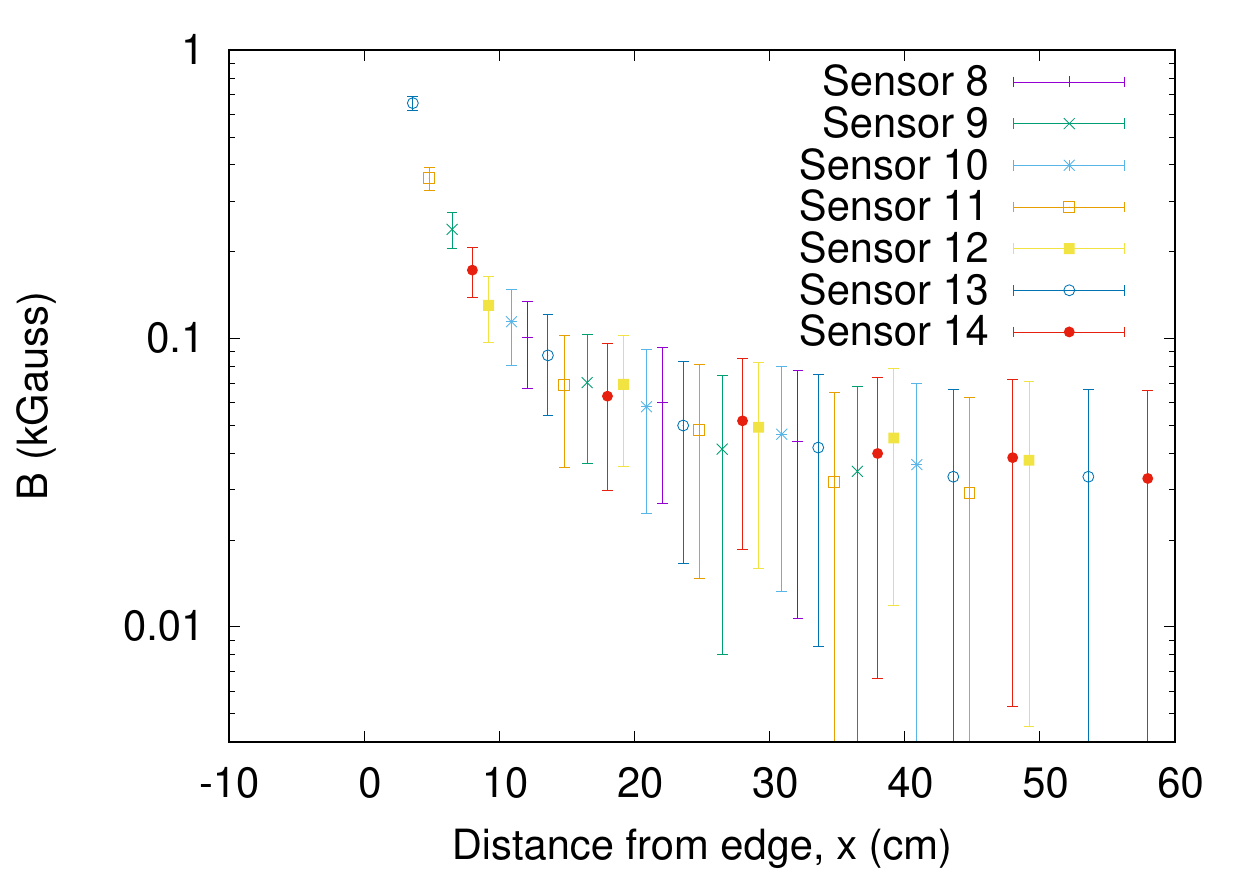}
\caption{Magnetic field in Gap 2 as a function of the distance $x$
from the edge of the iron plate. Negative values of $x$ indicate the
measurement is made inside {\sf mini-ICAL} while positive values indicate a
measurement in the air outside the detector. The figure on the right
(y-axis is in log scale) shows a close-up of the smaller fringe fields just outside the
detector every 10th point and only a few typical sensors have been
plotted for clarity.}
\label{fringe_field}
\end{figure}

Fig.~\ref{fringe_field} shows the magnetic field measured by the 14
sensors as a function of the distance $x$ from the detector edge. The
field is fairly uniform inside, dropping slightly towards the edge and
then dropping rapidly in the air outside the detector. The relatively
small fringe field in the air is also shown in close-up view in
Fig.~\ref{fringe_field} on a log scale so that the differences between
the field measured by different sensors can be seen. The measurements
from different sensors are consistent to within 0.01 kGauss. The errors have
been computed as per Eq.~\ref{eq:formula}. It can be seen that the
fringe field is about $B \sim 0.10 \pm 0.03$ kGauss at $x=10$ cm and
$B \sim 0.04 \pm 0.03$ kGauss at $x=50$ cm, which is at the limit of
sensitivity of the sensor.

\section{Discussion and future plans}
With a simple, cost-effective design and locally fabricated probes, a
measurement of the magnetic field was made in different locations in the
{\sf mini-ICAL} detector. This was the first such detailed measurement
of the magnetic field which will play a crucial role achieving the
physics goals of the main ICAL detector, including the determination of
the neutrino mass ordering/mass hierarchy.

The magnetic field in each layer of {\sf mini-ICAL} varies in both
magnitude and direction across the layer. Since there is provision for
measuring the magnetic field in three layers of {\sf mini-ICAL} (bottom,
middle and top) and each layer has 8 air gaps (including two which can
accommodate 2 Hall sensor PCBs side by side) to insert Hall probe PCBs,
the future plan is to make 30 such PCBs to make a precision measurement
of the magnetic field in different locations in {\sf mini-ICAL}.
An automated system will be developed for the data acquisition from the
Hall probe PCBs so that magnetic field from all the gaps can be measured and
recorded simultaneously.

Additionally, the magnetic field has currently been measured in the
linear, not saturation, region of the $B$-$H$ curve since the maximum
current that can be supplied to the coils is 500 A; this can generate a
field of $\sim 10.5$ kG. In the future we plan to measure the magnetic
field at saturation values of about 15 kG using a current of $\sim 900$ A by
installing a power supply which can provide currents of upto 1000 A. We
will also compare the measured and simulated magnetic fields that will
be an important input to determine a detailed magnetic field map for
both the {\sf mini-ICAL} and main ICAL detectors.

%\section{Appendix}
\appendix
\section{Calibration of Hall probe at 90$^{0}$ dipole magnet at ECR ion source facility.}

\begin{figure}[!tbp]
  \centering
  \begin{minipage}[b]{0.4\textwidth}
    \includegraphics[width=\textwidth]{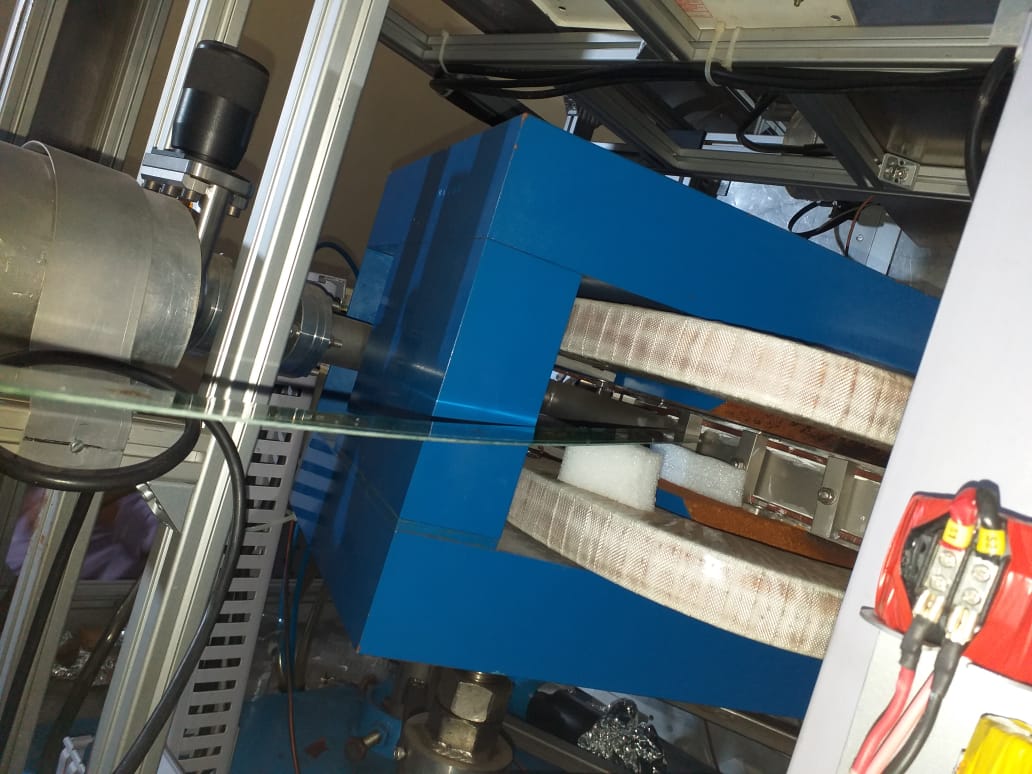}
    \caption{ECR Ion source magnet.}
    \label{ECR_ion_magnet}
  \end{minipage}
  \hfill
\begin{minipage}[b]{0.45\textwidth}
    \includegraphics[angle=0,width=\textwidth]{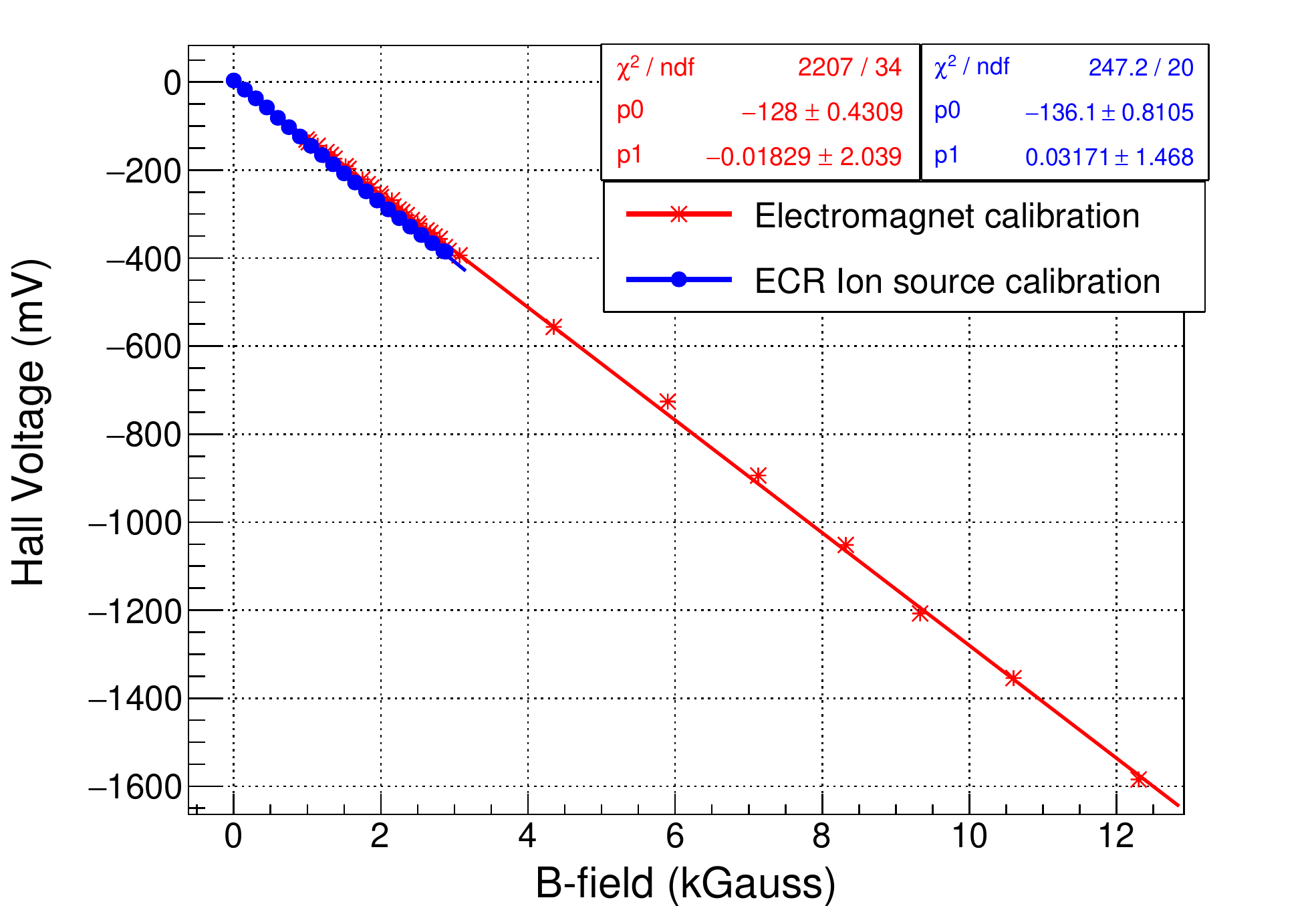}
    \caption{Comparison of the calibration done using electromagnet and ECR Ion source magnet.}
    \label{ECR_cal}
  \end{minipage}
\end{figure}

The Electron Cyclotron Resonance (ECR) lab in TIFR has a 90$^0$ bending
dipole electromagnet that is used to select ions having the same
mass/charge ($M/Q$) ratio. The maximum field that can be produced in it
is 3 kG. The system is controlled using LabVIEW software. The
photograph of the dipole magnet is shown in Fig-\ref{ECR_ion_magnet}.

The Hall probe PCB is placed in the middle of the two coils and oriented
such that the Hall probe is perpendicular to the magnetic field. The
field is increased in steps of 150 Gauss and corresponding Hall voltage
in the Hall sensor is noted (see Fig.~\ref{ECR_cal}). The Hall voltage vs
magnetic field data is plotted and fitted to a straight line for a field
upto 3 kG. The measurement is shown in Fig.~\ref{ECR_cal} along with the
calibration using the gap dipole electromanget discussed in
Section~\ref{sec:calib}. A reasonable consistency is seen at the lower
values of the field, $B \le 3$ kG, {\em i.e.}, in the overlapping regions
of measurement.

%\paragraph{Note added.} This is also a good position for notes added
%after the paper has been written.


\end{document}